\documentclass[preprint,12pt]{elsarticle}

\journal{Physics Letters B}

\usepackage{amsmath}
\usepackage{geometry}
\usepackage{graphicx}
\usepackage{lineno,hyperref}
\modulolinenumbers[5]
\usepackage{array}
\usepackage{microtype}
\usepackage{float}
\usepackage{amsfonts}
\usepackage{mathrsfs}
\usepackage{longtable}
\usepackage{xcolor}
\usepackage{epstopdf}
\hypersetup{colorlinks,citecolor=blue}
\usepackage{epsfig}
\usepackage{graphicx}
\usepackage{amsmath,amssymb}
\usepackage{epstopdf}
\usepackage{bm}
\usepackage{amsfonts}
\usepackage{lineno,hyperref}
\usepackage{dcolumn}

\begin{document}

\begin{frontmatter}

\title{On the reconstruction of the  Rotation Curve for Milky Way and its spacetime implications: a Machine Learning approach}

\author[label1]{Aritra Sanyal}\ead{aritrasanyal1@gmail.com}
\author[label2]{Swapan Das}\ead{swdas@arden.ac.uk}
\author[label1]{Farook Rahaman}\ead{rahaman@associates.iucaa.in}
\author[label3]{Saibal Ray\footnote{Corresponding author}}\ead{saibal.ray@gla.ac.in}

\address[label1]{Department of Mathematics, Jadavpur University, Kolkata 700032, West Bengal, India}
\address[label2]{Centre for Academic Persistence, Arden University, London, United Kingdom} 
\address[label3]{Centre for Cosmology, Astrophysics and Space Science (CCASS), GLA University, Mathura 281406, Uttar Pradesh, India}

\begin{abstract}
We propose a machine learning-assisted analytical reconstruction of the Milky Way rotation curve and discuss the implications thereof in a relativistic spacetime context. The rotation curve is reconstructed using 73 observational data points over the range 0.1 to 95.56 kpc, and we compare the performances of Ridge regression, LASSO regression, and feed-forward neural networks on a physically motivated functional basis. Ridge regression yields the most stable prediction with $R^2 = 0.9824 \pm 0.0064$, $RMSE = 3.75~km~s^{-1}$, with the additional advantage of retaining analytical interpretability. We embed the reconstructed velocity profile into a static spherically symmetric space time, which facilitates the determination of the redshift function and the mass function using Einstein’s equations. We verify that all energy conditions are satisfied, the sound speed is subluminal, the circular orbits are stable, and the gravitational energy is negative, confirming the attractive nature of the gravitational force. This framework constitutes a statistically validated, data-driven alternative to conventional dark matter halo models, establishing a clear connection between the kinematical observables and the relativistic space time geometry.
\end{abstract}

\begin{keyword}
rotation curve; machine learning; galactic dynamics; spacetime geometry; dark matter
\end{keyword}

\end{frontmatter}

\section{Introduction}

Rotation curves of spiral galaxies constitute one of the most robust observational probes of gravitational dynamics on galactic scales \cite{Rubin1980,Sofue2001}. The persistent deviation of observed rotational velocities from the Keplerian expectation at large radii has long indicated the presence of additional gravitational effects beyond those attributable to luminous matter alone \cite{Bosma1981}. In the case of the Milky Way, high-precision measurements spanning from the inner bulge to the extended halo reveal a complex radial velocity structure, characterized by a rapid rise in the inner regions, a near-flat intermediate regime, and a gradual decline in the outer halo \cite{Sofue2009,Eilers2019}. Such behavior cannot be adequately captured by simple Newtonian models or overly restrictive phenomenological prescriptions.

Traditional analyses of galactic rotation curves typically rely on specific dark matter halo profiles, such as the Navarro--Frenk--White (NFW) or Burkert models, whose parameters are inferred through fitting procedures \cite{Navarro1997,Burkert1995}. While these models are physically motivated and widely used, they impose predetermined functional forms on the data, which can bias the inferred mass distribution and limit the flexibility required to describe the full complexity of the Milky Way rotation curve \cite{deBlok2010}. Moreover, parametric halo models are not naturally suited for direct embedding into relativistic spacetime frameworks, where the metric functions must be smooth, differentiable, and physically consistent over the entire radial domain \cite{Rahaman2012}.

An alternative and complementary strategy is to reconstruct the rotation curve directly from observational data in a model-independent yet physically interpretable manner. Recent advances in machine learning have demonstrated significant potential in astrophysical data analysis, particularly for regression problems involving noisy and heterogeneous datasets \cite{Ball2010,Baron2019}. However, many machine-learning approaches emphasize predictive accuracy at the expense of analytical transparency, thereby limiting their applicability to theoretical modeling within general relativity \cite{Ntampaka2019}.

In this work, we adopt a machine-learning-guided regression framework designed explicitly to preserve analytical interpretability. By employing a physically motivated basis of functions and systematically comparing Ridge regression, LASSO regression, and feed-forward neural network models, we extract a smooth and explicit analytical expression for the Milky Way rotation curve directly from observational data \cite{Hoerl1970,Tibshirani1996}. The emphasis is not on black-box prediction, but on constructing a velocity profile that can be meaningfully embedded into a relativistic spacetime geometry.

The reconstructed velocity profile is subsequently incorporated into a static and spherically symmetric spacetime, allowing the redshift function and mass function to be determined self-consistently from Einstein’s field equations \cite{Einstein1915,Lake2004}. This procedure establishes a direct connection between galactic kinematics and spacetime geometry, enabling a detailed investigation of the effective matter content supporting the galactic halo. We analyze the resulting energy density, anisotropic pressure, equation of state, classical energy conditions, causality constraints, and the stability of circular orbits \cite{Visser1996,Herrera1997,Wang2026energy,Wang2026families}, thereby assessing the physical viability of the reconstructed spacetime.

This study builds upon our earlier observationally guided construction of redshift functions derived from statistically validated Milky Way rotation curves \cite{rahaman2026observationally,Sanyal2026mnras}. In that work, a polynomial reconstruction of the azimuthal velocity profile was employed to infer a physically viable spacetime geometry and to examine energy conditions, causality, orbital stability, and gravitational redshift effects. The present analysis advances this framework by introducing a machine-learning-assisted reconstruction strategy, which enhances flexibility and fidelity while retaining full analytical control and physical interpretability.

Overall, this work provides a transparent and data-driven bridge between observed galactic rotation curves and relativistic spacetime geometry. By avoiding restrictive \emph{a priori} assumptions on the underlying mass distribution while preserving analytical tractability, the proposed framework offers a complementary alternative to conventional parametric halo models and contributes to a deeper understanding of dark-matter-like gravitational effects in galactic systems.

Complementary studies of galactic gravitational geometry reconstructed from photometric and kinematic profiles, such as that of NGC 7331~\cite{Sanyal2026mnras}, further motivate the present data-driven approach for the Milky Way. The outline of the investigation is as follows: in section 2, we present the line element and relevant field equations for the modeling whereas Machine Learning reconstruction of the rotation curve has been provided in section 3. We have performed physical analysis (Mass, Density and Pressure (5.1), Energy conditions (5.2),  Equation of State (5.3), and velocity of sound (5.4). In section 6, stability of circular orbits are discussed whereas in section 7, we have analysed attractive gravity from effective anisotropic stress. All the plots shown in the manuscript are explained in detailed based on their physical features and viability are put in section 8. In section 9, we discuss and make a few concluding remarks on our study.

\section{Line element and Field Equations}

We consider a general static and spherically symmetric spacetime characterized by the line element~\cite{oppenheimer1939,delgaty1998,lake2003all,Wang2026energy,Wang2026families}:
\begin{equation}
    ds^2 = -f(r)\, dt^2 + \frac{dr^2}{1 - \frac{2 m(r)}{r}} + r^2 d\Omega^2,
\end{equation}
where \( f(r) \) is the redshift function encoding the gravitational potential, and \( m(r) \) is the mass function representing the total mass enclosed within the radial coordinate \( r \)~\cite{mak2003,chandrasekhar1935,bekenstein1971}.

To describe the matter distribution that supports this geometry, we adopt an anisotropic fluid model in which the radial pressure vanishes. The energy-momentum tensor for such a configuration is given by~\cite{mak2003}
\begin{equation}
    T^0_0 = -\rho(r), \quad T^1_1 = 0, \quad T^2_2 = T^3_3 = p(r),
\end{equation}
where \( \rho(r) \) denotes the energy density, and \( p(r) \) is the tangential pressure.

Substituting this form of the energy-momentum tensor into Einstein's field equations leads to the following relations, consistent with the general treatments of static spherically symmetric matter distributions discussed in~\cite{Wang2026energy,Wang2026families}. The first one is the differential equation for the mass function~\cite{MS1964}:
\begin{equation} \label{efe1}
    m'(r) = 4 \pi r^2 \rho(r),
\end{equation}
which describes how the mass increases with radius depending on the energy density profile~\cite{oppenheimer1939,tolman1939}.

The second equation arises from the \( tt \)-component of the field equations and governs the redshift function
\begin{equation} \label{efe2}
    \frac{f'(r)}{f(r)} = \frac{2m(r)}{r^2 - 2r\, m(r)},
\end{equation}
showing how \( f(r) \) is influenced by the geometry and the mass distribution~\cite{bekenstein1971,mak2003}.

Lastly, the \( \theta\theta \) or \( \phi\phi \)-components provide the expression for tangential pressure in terms of \( \rho(r) \) and \( m(r) \):
\begin{equation} \label{efe3}
    p(r) = \frac{1}{2} {r\, \rho(r)} \left[\frac{m(r)}{r^2 - 2r\, m(r)}\right].
\end{equation}

These three coupled equations form the foundation of our model and will be used to investigate the physical viability of the spacetime once an explicit form for \( f(r) \) is known from observational data.

\section{Machine Learning Methodology}

\subsection{Machine Learning Modeling of the Rotation Curve}

The Milky Way rotation curve dataset consists of $73$ observational points spanning galactocentric radii from $0.100$ to $95.560$~kpc, with azimuthal velocities in the range $145$--$251$~km~s$^{-1}$ (vide Appendix A).

To model the velocity profile, we construct a library of physically motivated basis functions incorporating power-law terms, inverse-radius contributions, logarithmic dependence, and exponential decay and saturation terms. This basis is designed to capture both inner-galaxy behavior and outer-halo trends.

\subsubsection{Basis Construction and Reprocessing}

The regression model was built upon a set of basis functions inspired by physics, which included power law, inverse radius, logarithm, exponential decay, and exponential saturation terms. These basis functions were chosen based on their potential to reproduce the dominant characteristics of the Milky Way rotation curve, namely its steep rise in the inner region, flat shape in the intermediate region, and slow decline in the outer region. 
Radial Distance values were used as input feature for this regression model. Before applying the regression model, data was standardized for numerical stability and comparison purposes. 

\subsubsection{Model Training and Selection}

The models considered were Ridge regression, LASSO regression, and a feed-forward neural network model. For regression models, parameters were chosen through cross-validation to prevent overfitting and maintain model accuracy. A $k$-fold cross-validation method was adopted for evaluating model performance and ensuring robustness of model results. The neural network model architecture was composed of fully connected layers and utilized standard backpropagation and mean squared error as the performance metric. Model performance was evaluated by using the coefficient of determination ($R^2$), root mean square error, and reduced chi-square values.

\subsubsection{Uncertainty Estimation}

The uncertainty bands plotted in the rotation curves were based on the variability of model predictions made during cross-validation. The $1\sigma$ and $2\sigma$ confidence intervals represent the variability of predicted velocities across different validation sets, which is an estimate of model uncertainty. Observational uncertainties were also considered in velocity measurements for model performance evaluation.\\ 

Altogether three regression techniques are applied to the same dataset and basis set: (i) Ridge regression, (ii) LASSO regression, and (iii) a feed-forward neural network.

Ridge regression employs $\ell_2$ regularization and yields the most accurate and stable reconstruction, with a coefficient of determination $R^2 = 0.9824$ and a root
mean square error of $3.75$~km~s$^{-1}$ (vide Appendix B).

LASSO regression enforces sparsity and retains only a small subset of basis functions, but at the expense of significantly reduced accuracy.

A neural network with two hidden layers provides a nonlinear benchmark but remains inferior to the Ridge-based model in both accuracy and interpretability.

Given its superior performance and analytical transparency, the Ridge regression solution is adopted as the final reconstructed velocity law.

\begin{figure}[h]
\centering
\includegraphics[width=0.95\linewidth]{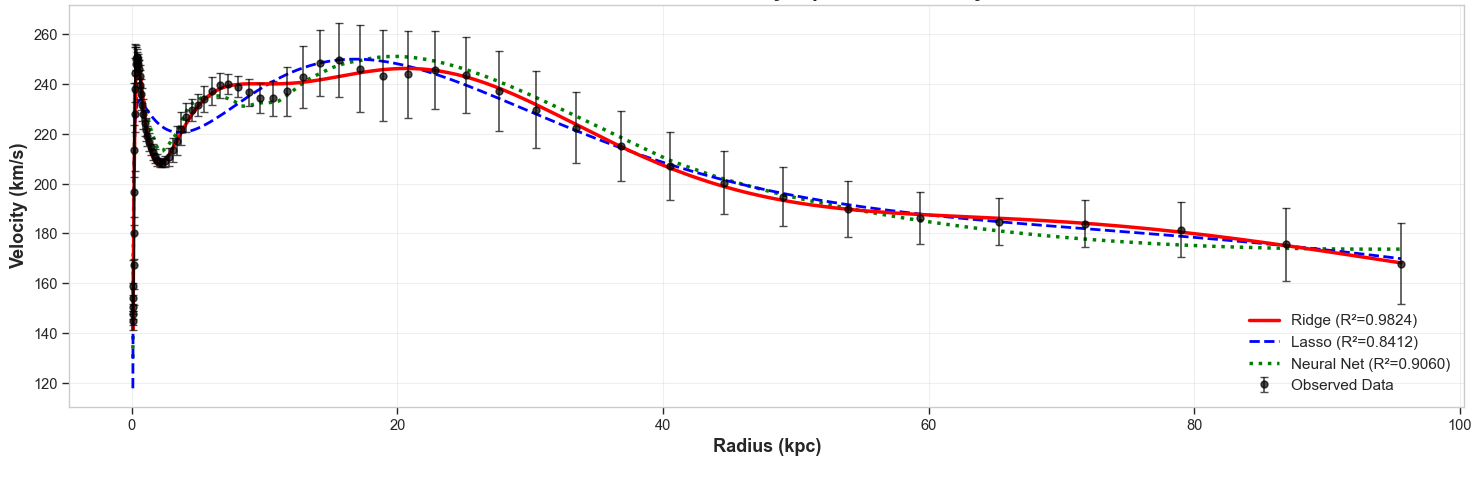}
\caption{Observed Milky Way rotation curve data and the reconstructed velocity profile
obtained from Ridge regression.}
\label{fig:mlvelocity}
\end{figure}

The resulting azimuthal velocity profile is expressed as
\begin{equation}
\begin{split}
v(r) &= 52.966\,r - 393.881\,r^{1/2} - 88.773\,r^{-1} + 3.968\,r^{-2}
 - 99.023\,e^{-r/2} - 16.888\,e^{-r/5} \\
&\quad - 3.897\,e^{-r/10} - 1.920\,e^{-r/20}
 + 89.013\,(1-e^{-r/3}) + 4.931\,(1-e^{-r/8}) \\
&\quad + 2.588\,(1-e^{-r/15}) + 1.507\,(1-e^{-r/25})
 - 133.453\,\log_{10}(r+1) - 307.286\,\ln(r+1) \\
&\quad + 325.930\,r\,e^{-r/5} + 38.245\,r\,e^{-r/10}
 + 40.713\,r\,e^{-r/20}
 + 11.913\,r^2\,e^{-r/10} \\
&\quad - 3.446\,r^2\,e^{-r/15}
 + 2.630\,r^2\,e^{-r/25}
 - 424.416\,r^{1/2}\,e^{-r/10},
\end{split}
\label{eq:v_ml}
\end{equation}
where the radial coordinate $r$ is measured in kilo-parsecs (kpc) and the azimuthal velocity
$v(r)$ is expressed in km\,s$^{-1}$.

\subsection{Computing redshift function from observed velocity data}

Using the unified Milky Way rotation curve compiled by \cite{Sofue2020MilkyWayRC}, we model
the azimuthal velocity profile using a machine-learning-guided regression approach which has been provided in Eq. (\ref{eq:v_ml}). 

In a static and spherically symmetric spacetime, the redshift function $f(r)$ is related to
the azimuthal velocity of test particles in circular orbits through the relativistic relation
\begin{equation}
\frac{\mathrm{d}f}{\mathrm{d}r} = \frac{v^{2}(r)}{r}.
\label{eq:MW_redshift_derivative}
\end{equation}

Integrating Eq.~\eqref{eq:MW_redshift_derivative}, we obtain
\begin{equation}
f(r) = \int \frac{v^{2}(r)}{r}\,\mathrm{d}r.
\label{eq:MW_redshift_integral}
\end{equation}

Substituting the velocity expression from Eq.~\eqref{eq:v_ml} and performing the integration analytically, the explicit form of the redshift function is obtained.

\section{Comparison with other models}

We compare the predictive performance of the Ridge regression model with physically motivated dark matter halo models, namely the Navarro--Frenk--White (NFW) and Burkert profiles. The evaluation is based on cross-validation statistics, standard regression accuracy metrics, and goodness-of-fit diagnostics.

\subsection{Overall Performance with Uncertainties}

The Ridge regression model exhibits excellent predictive capability, achieving a cross-validated coefficient of determination of
\[
R^2 = 0.9824 \pm 0.0064,
\]
along with a low root-mean-square error (RMSE) of $3.75~\mathrm{km\,s^{-1}}$. The associated prediction uncertainty is
\[
\sigma = 8.71 \pm 19.91~\mathrm{km\,s^{-1}},
\]
indicating strong consistency across the dataset.

In contrast, the NFW halo model yields a significantly lower cross-validated performance with
\[
R^2 = 0.6430 \pm 3.32,
\]
and a substantially larger RMSE of $16.88~\mathrm{km\,s^{-1}}$. The corresponding prediction uncertainty is
\[
\sigma = 4.95 \pm 2.20~\mathrm{km\,s^{-1}},
\]
reflecting notable dispersion and reduced predictive reliability.

The Burkert profile performs moderately better than the NFW model, with a cross-validated coefficient of determination
\[
R^2 = 0.7915 \pm 5.09,
\]
and an RMSE of $12.90~\mathrm{km\,s^{-1}}$. Its prediction uncertainty,
\[
\sigma = 3.31 \pm 1.92~\mathrm{km\,s^{-1}},
\]
suggests improved stability relative to the NFW profile, though still inferior to the Ridge model.

\subsection{Model Accuracy Metrics}

A detailed comparison of standard accuracy metrics further highlights the superiority of the Ridge regression approach. The Ridge model attains
\[
R^2 = 0.9824,\quad
\mathrm{RMSE} = 3.75 \pm 1.17~\mathrm{km\,s^{-1}},\quad
\mathrm{MAE} = 2.72~\mathrm{km\,s^{-1}},
\]
with a reduced chi-square value of
\[
\chi^2_{\mathrm{red}} = 0.350,
\]
indicating an excellent fit to the observational data.

By comparison, the NFW model shows poor agreement with the data, characterized by
\[
R^2 = 0.6430,\quad
\mathrm{RMSE} = 16.88 \pm 3.84~\mathrm{km\,s^{-1}},\quad
\mathrm{MAE} = 13.87~\mathrm{km\,s^{-1}},
\]
and a large reduced chi-square of
\[
\chi^2_{\mathrm{red}} = 11.05,
\]
suggesting a statistically unacceptable fit.

The Burkert profile yields marginal performance, with
\[
R^2 = 0.7915,\quad
\mathrm{RMSE} = 12.90 \pm 5.31~\mathrm{km\,s^{-1}},\quad
\mathrm{MAE} = 10.68~\mathrm{km\,s^{-1}},
\]
and a reduced chi-square value of
\[
\chi^2_{\mathrm{red}} = 7.44.
\]
While this represents an improvement over the NFW model, it remains significantly less accurate than the Ridge regression model.

\begin{figure}
    \centering
    \includegraphics[width=1\linewidth]{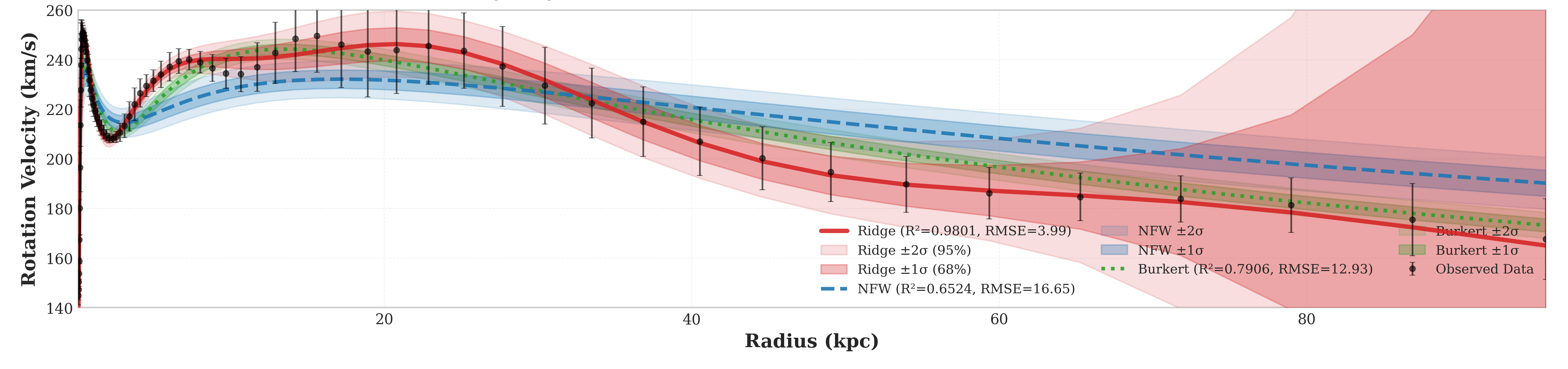}
    \caption{Combined comparison of the Milky Way rotation curve with Ridge regression, Navarro--Frenk--White (NFW), and Burkert model predictions, showing mean velocity profiles with $1\sigma$ and $2\sigma$ confidence intervals overlaid on observational data.}

    \label{fig:placeholder}
\end{figure}
\begin{figure}
    \centering
    \includegraphics[width=1\linewidth]{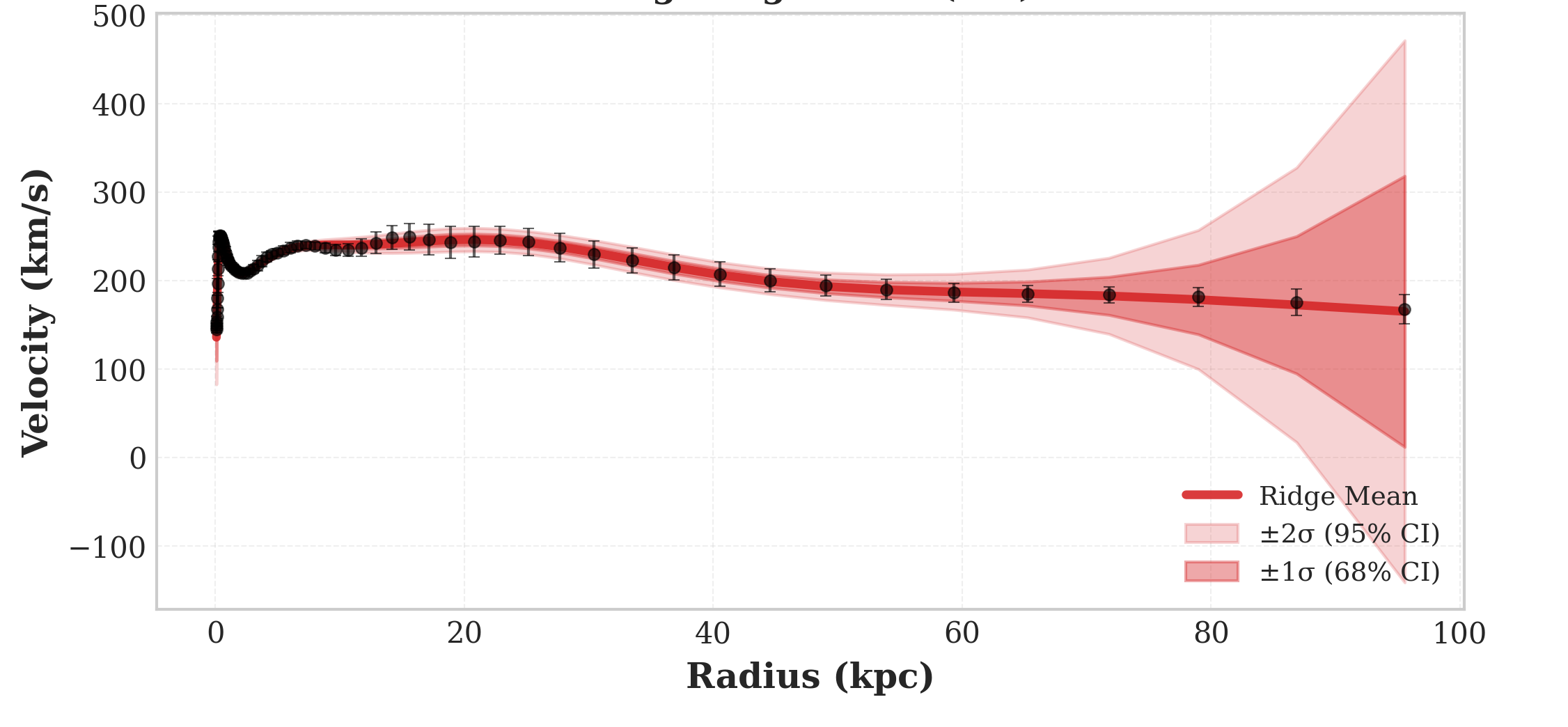}
    \caption{Ridge regression (machine learning) prediction of the Milky Way rotation curve, showing the mean velocity profile with $1\sigma$ (68\%) and $2\sigma$ (95\%) confidence intervals overlaid on observational data.}

    \label{fig:placeholder}
\end{figure}
\begin{figure}
    \centering
    \includegraphics[width=1\linewidth]{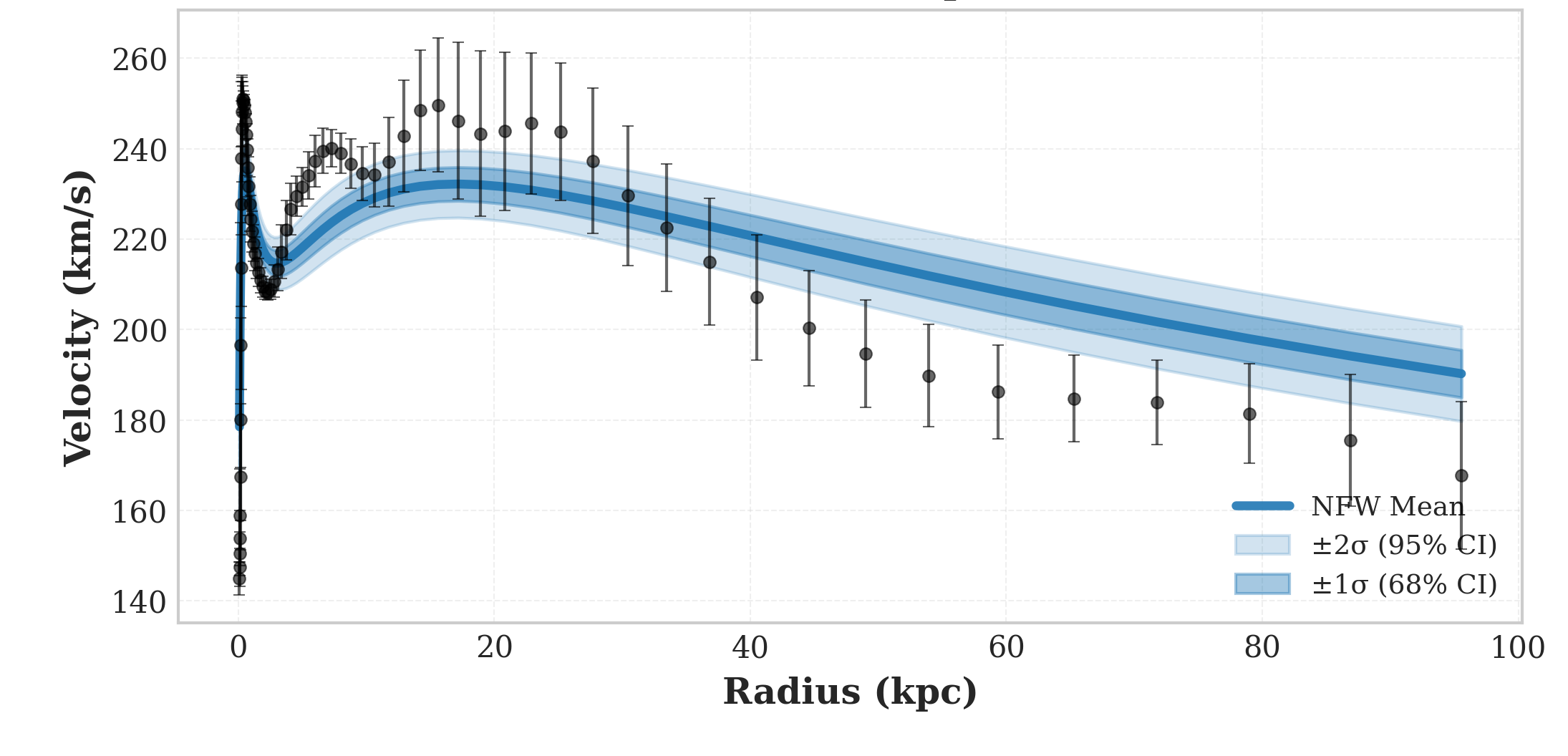}
    \caption{Rotation curve fit using the Navarro--Frenk--White (NFW) multi-component halo model, illustrating the mean prediction and associated $1\sigma$ and $2\sigma$ uncertainty bands compared with observations.}

    \label{fig:placeholder}
\end{figure}
\begin{figure}
    \centering
    \includegraphics[width=1\linewidth]{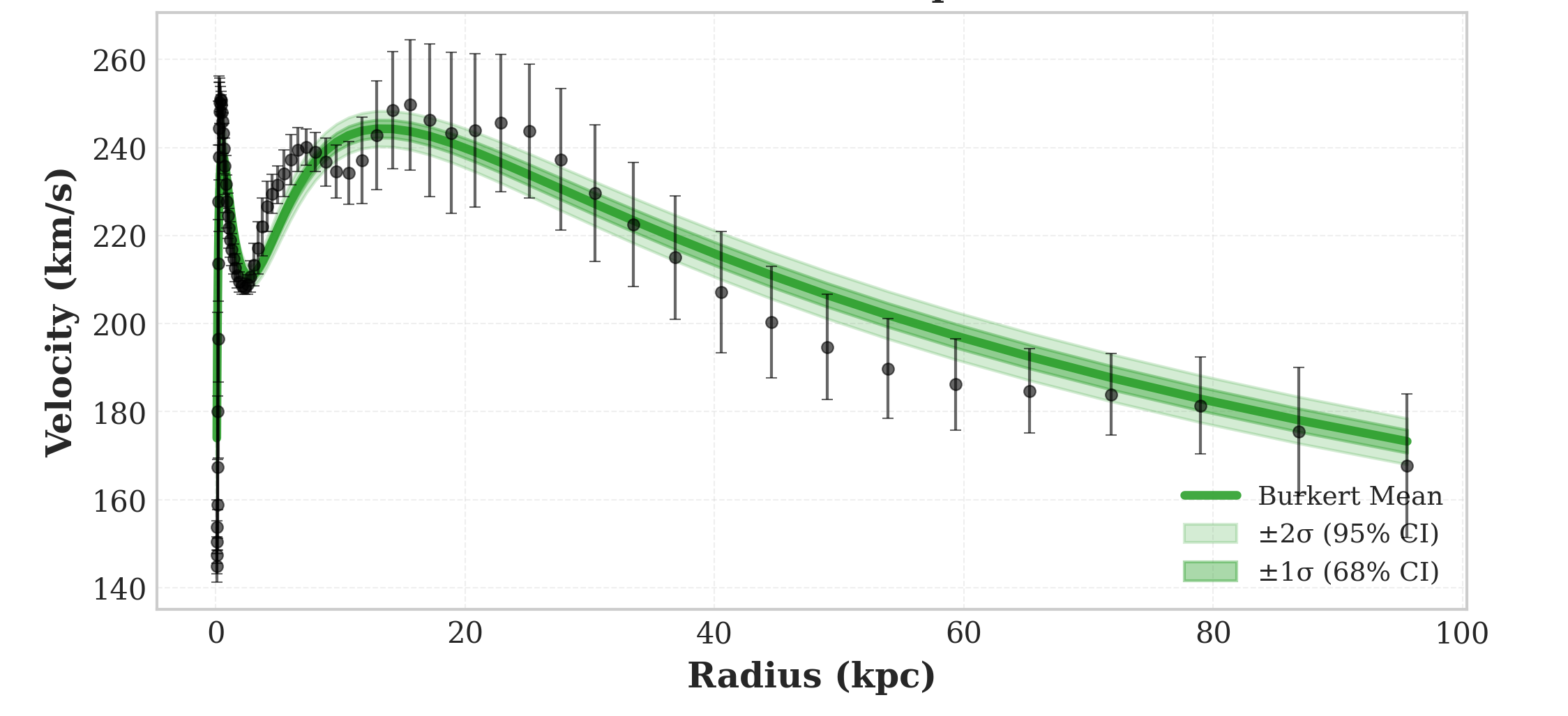}
    \caption{Burkert multi-component halo model fit to the Milky Way rotation curve, where the cored density profile yields improved agreement relative to the NFW model, with shaded confidence intervals indicating prediction uncertainty.}

    \label{fig:placeholder}
\end{figure}
\begin{figure}
    \centering
    \includegraphics[width=1\linewidth]{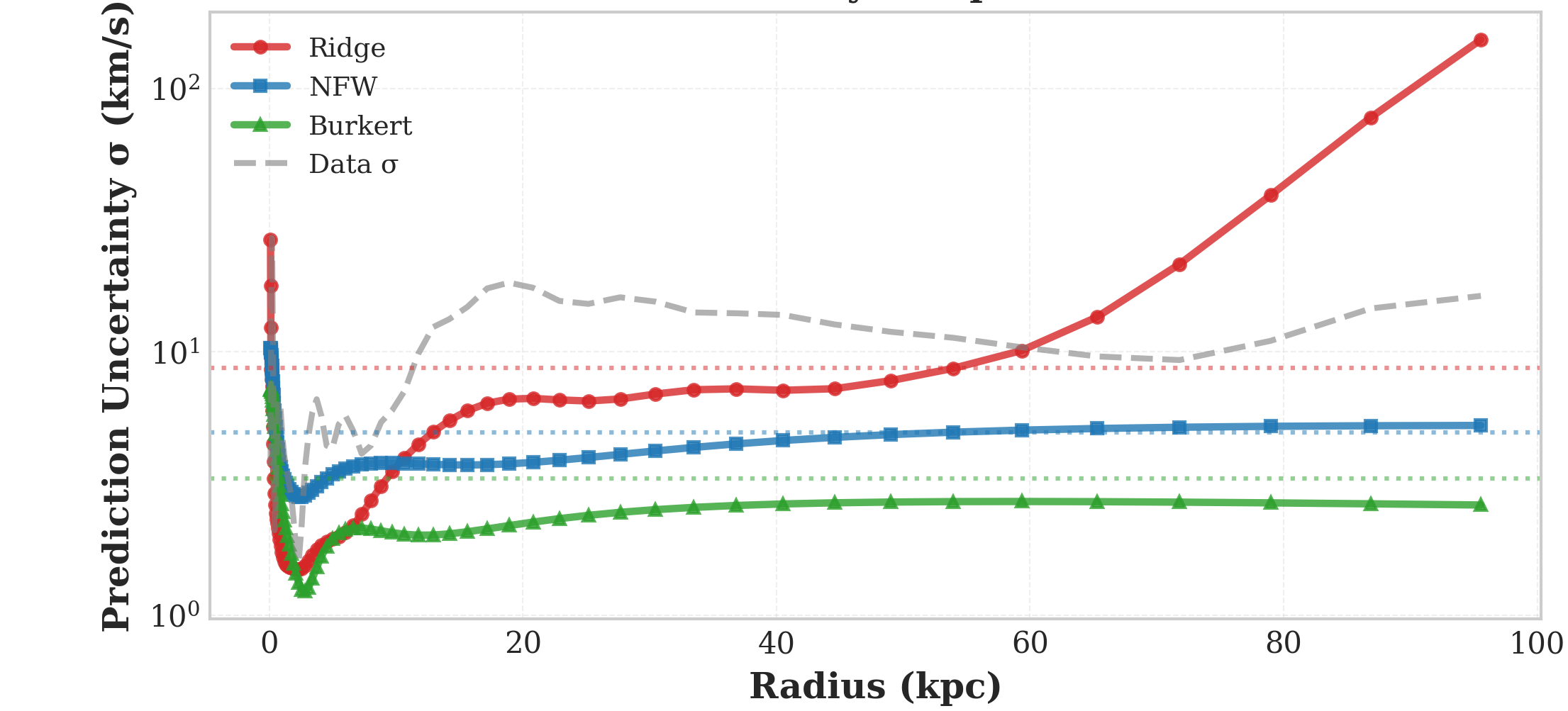}
    \caption{Radial dependence of the prediction uncertainty $\sigma$ for the Ridge regression, NFW, and Burkert models, shown alongside the observational uncertainty for comparison.}

    \label{fig:placeholder}
\end{figure}

\section{Physical analysis}

\subsection{Mass, Density and Pressure}

From the Einstein field equation \eqref{efe2}, we get the expression for $m(r)$ as
\begin{equation}
    m(r)=\frac{r^2 f'(r)}{2f(r)+2rf'(r)}.
\end{equation}

\begin{figure}
    \centering
    \includegraphics[width=1\linewidth]{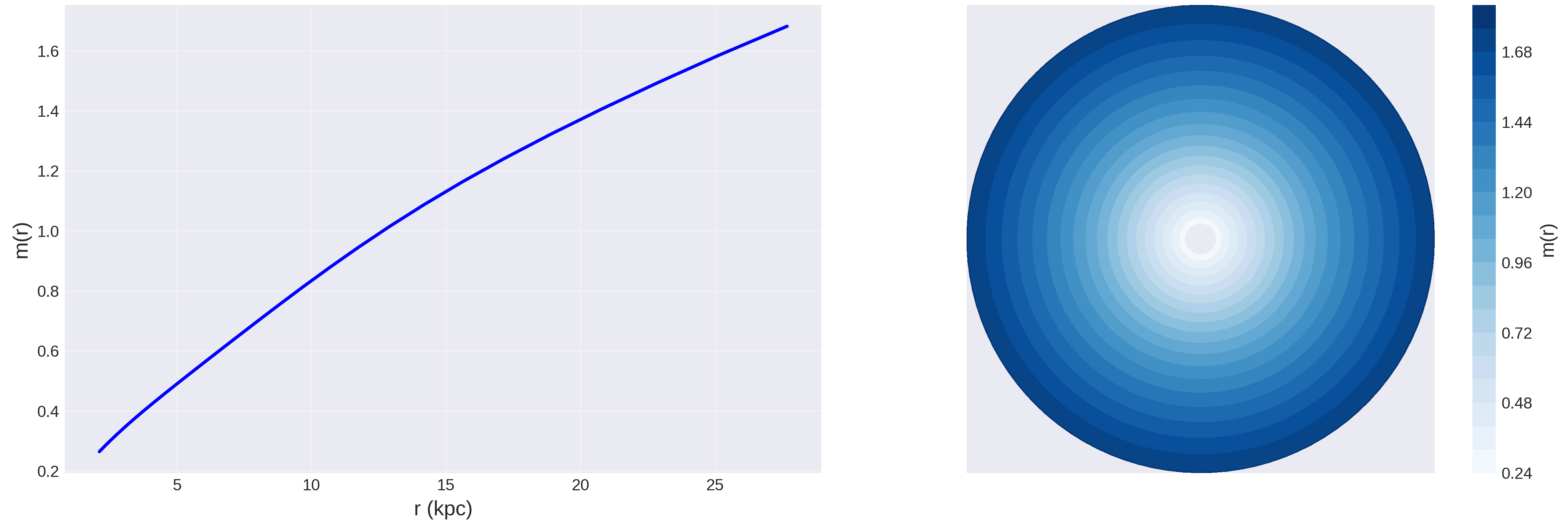}
     \caption{Plot for $m(r)$ vs $r$ for radial values and its corresponding contour plot}
    \label{fig:placeholder}
\end{figure}

\begin{figure}
    \centering
    \includegraphics[width=1\linewidth]{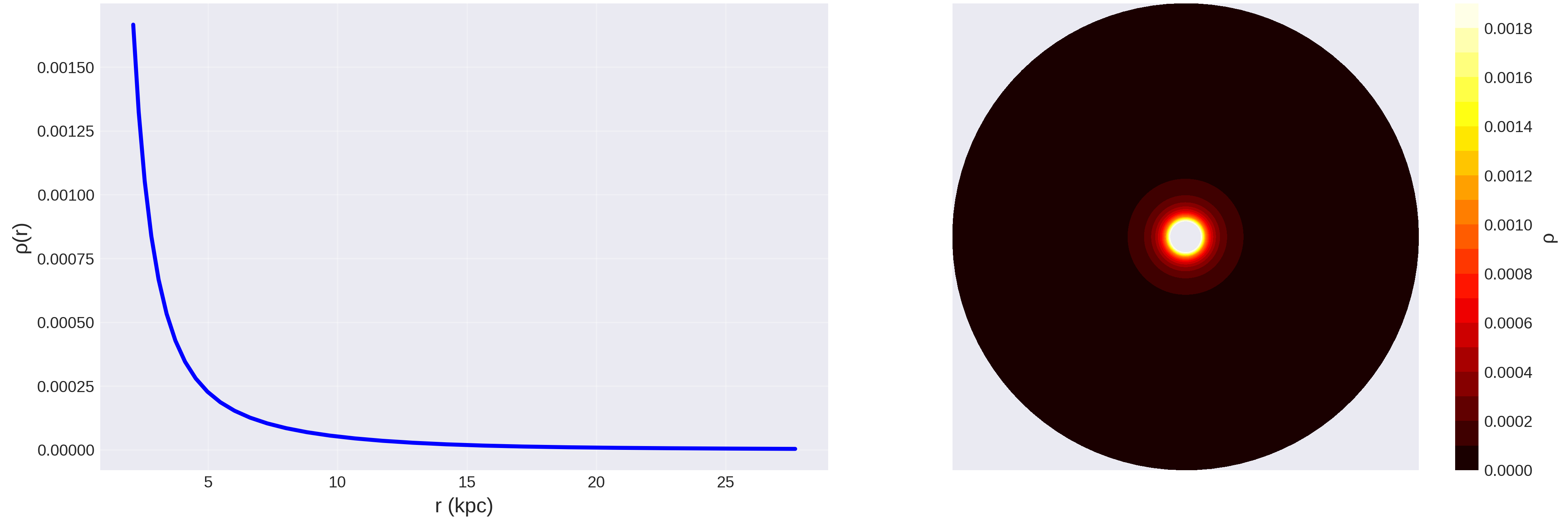}
    \caption{Plot for $\rho(r)$ vs $r$ and its corresponding contour plot }
    \label{fig:placeholder}
\end{figure}

Using the mass function given in Eq. (\ref{eq:MW_redshift_integral}) together with the Einstein equation (\ref{efe2}), the energy density can be written as
\begin{equation}
    \rho(r) = \frac{1}{4\pi r^2} \left[m'(r)\right].
\end{equation}

Further, using $m(r)$ we get the expression of $\rho$ and correspondingly, using Eq. \eqref{efe3}, we obtain $p$.

\begin{figure}
    \centering
    \includegraphics[width=1\linewidth]{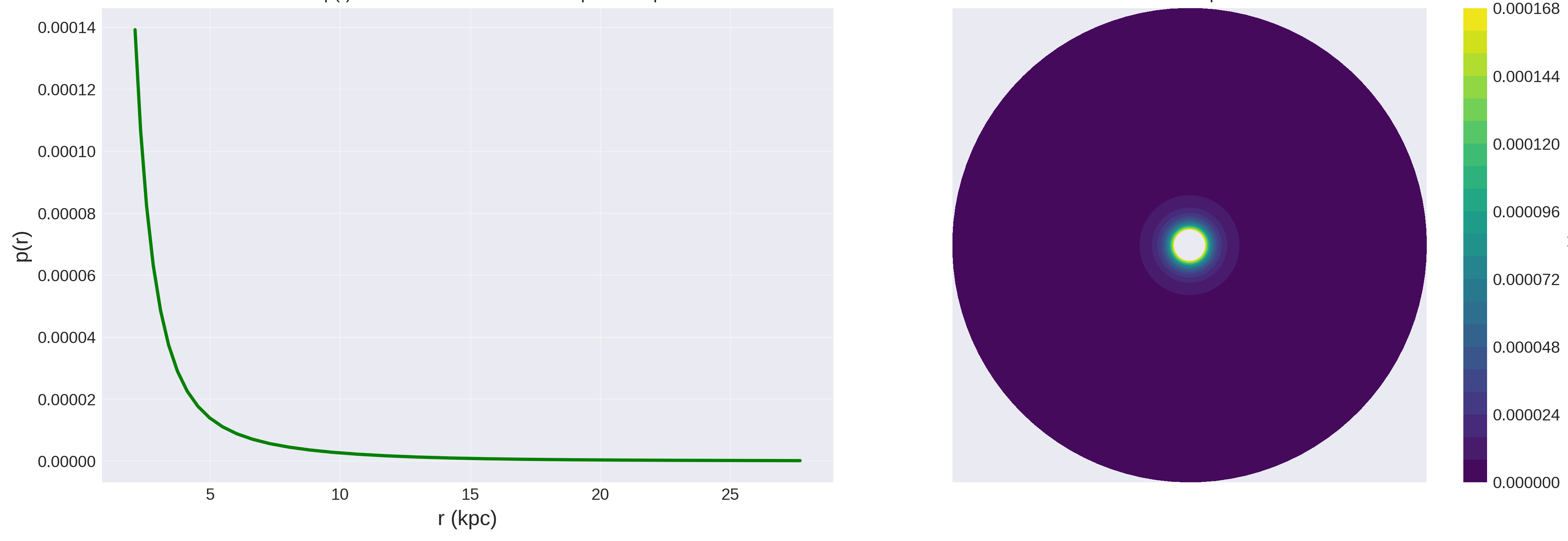}
    \caption{{Plot for $p(r)$ vs $r$ for radial values  and its corresponding contour plot}}
    \label{fig:placeholder}
\end{figure}

\subsection{Energy conditions}

The energy conditions are useful criteria to test physical viability of the effective matter distribution associated to the spacetime geometry \cite{Wang2026energy,Wang2026families}. In the case of the anisotropic fluid considered in this paper, where we have vanishing radial pressure $p_r = 0$ and tangential pressure $p_t = p$, the situation is as follows:

$\bullet$ Null Energy Condition (NEC) : $\rho + p_r \geq 0$, $\rho + p_t \geq 0$. 

$\bullet$ Weak Energy Condition (WEC) : $\rho \geq 0$, $\rho + p_r \geq 0$, $\rho + p_t \geq 0$.

$\bullet$ Strong Energy Condition (SEC) : $\rho + p_r \geq 0$, $\rho + p_t \geq 0$, $\rho + p_r + 2 p_t \geq 0$.

$\bullet$ Dominant Energy Condition (DEC) : $\rho \geq |p_r|$, $\rho \geq |p_t|$.\\

In our model, since the radial pressure vanishes, so the energy conditions simplify to\\

NEC: $\rho + p \geq 0$. 

WEC: $\rho \geq 0$, $\rho + p \geq 0$.

SEC: $\rho + p \geq 0$, $\rho + 2 p \geq 0$.

DEC:  $\rho \geq |p|$.
\newline

The results presented here demonstrate that all the above inequalities are well satisfied in the radial range under consideration. Thus, the effective matter content remains physically acceptable, and no violation of classical energy conditions is necessary to reproduce the observed galactic rotation properties.

\subsection{Equation of State (EOS)}

Next we take the EOS $p=\omega \rho$ ($\implies\omega=\frac{p}{\rho}$). From the calculated values of $\rho(r)$ and $p(r)$, we get $\omega$.

\begin{figure}
    \centering
    \includegraphics[width=1\linewidth]{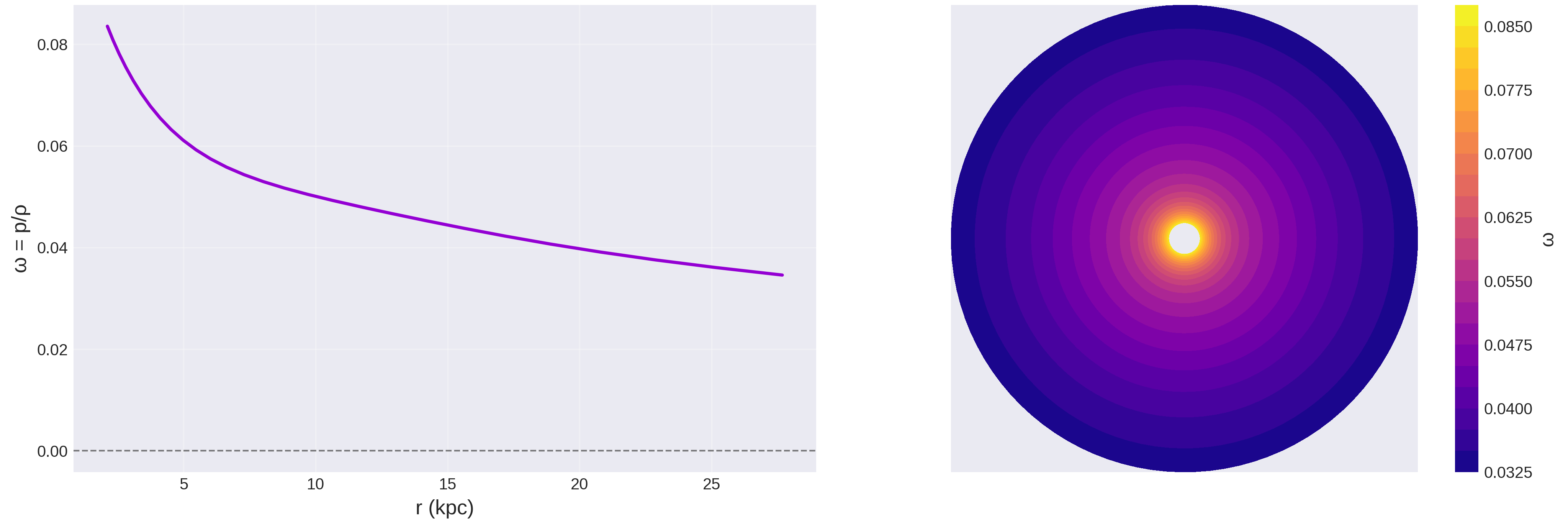}
    \caption{Plot for $\omega$ vs $r$ for $r$-values and its corresponding contour plot}
    \label{fig:placeholder}
\end{figure}

\begin{figure}
    \centering
    \includegraphics[width=1\linewidth]{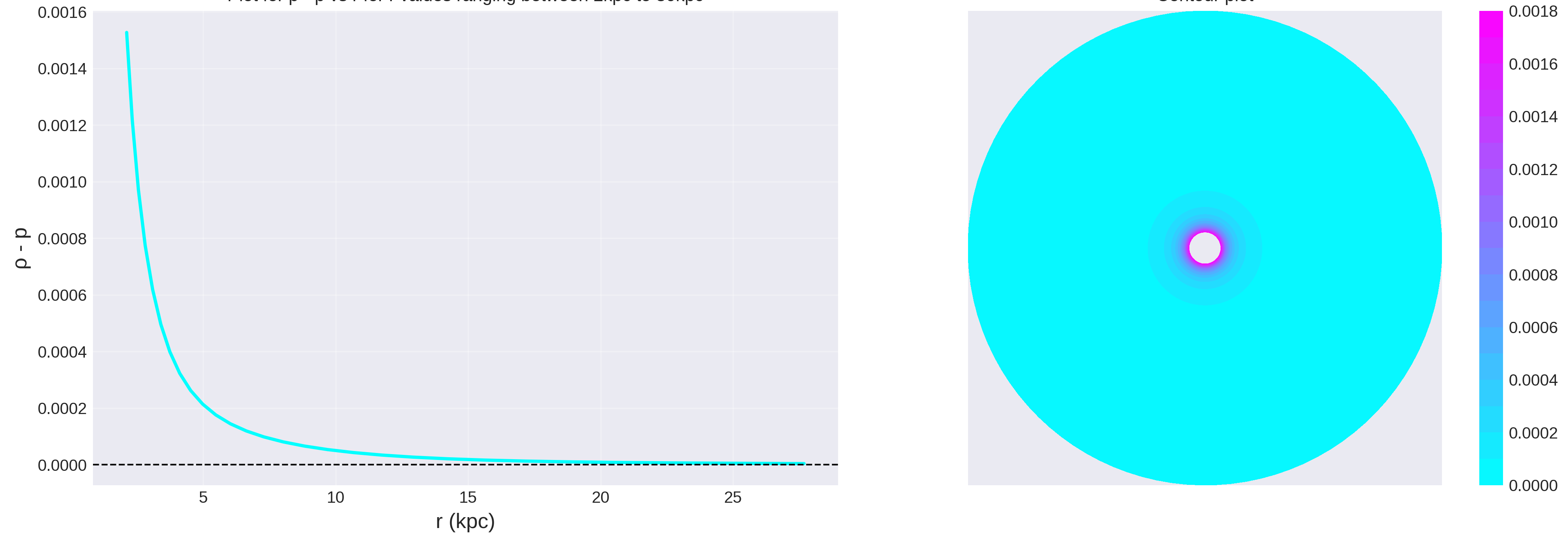}
    \caption{Plot for $\rho-p$ vs $r$ for $r$ values and its corresponding contour plot}
    \label{fig:placeholder}
\end{figure}
\begin{figure}
    \centering
    \includegraphics[width=1\linewidth]{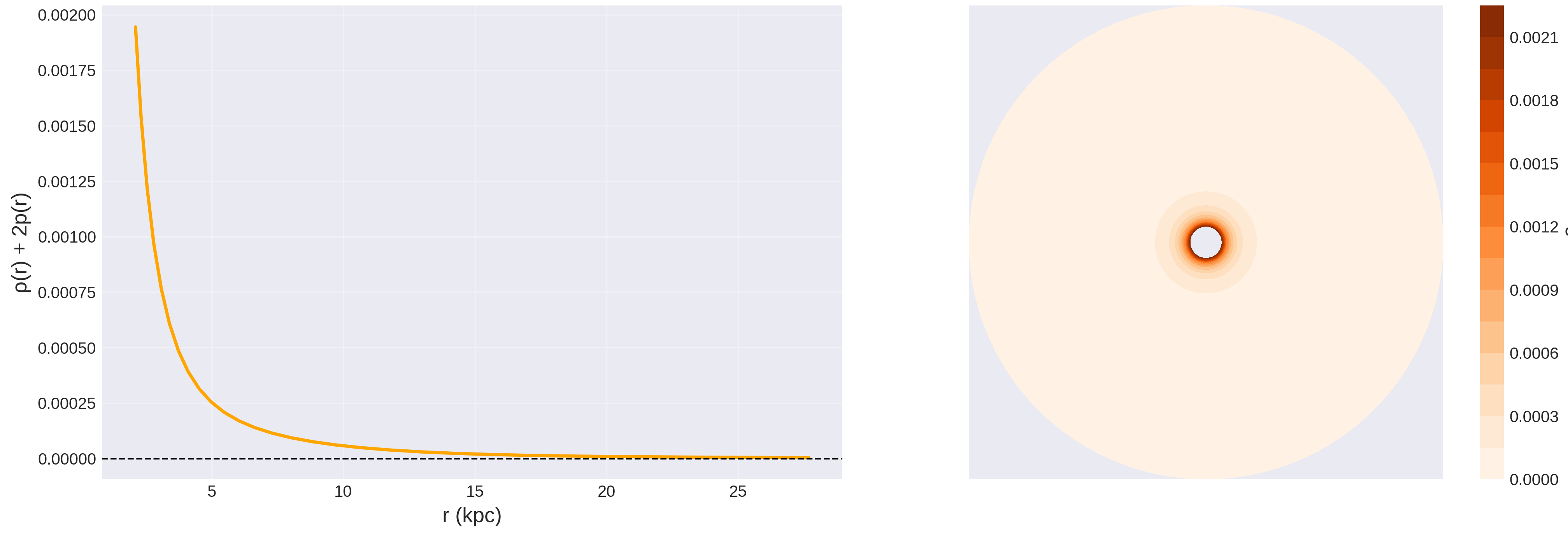}
     \caption{Plot for $\rho(r)+2p(r)$ vs $r$ for radial values and its corresponding contour plot}
    \label{fig:placeholder}
\end{figure}
\begin{figure}
    \centering
    \includegraphics[width=1\linewidth]{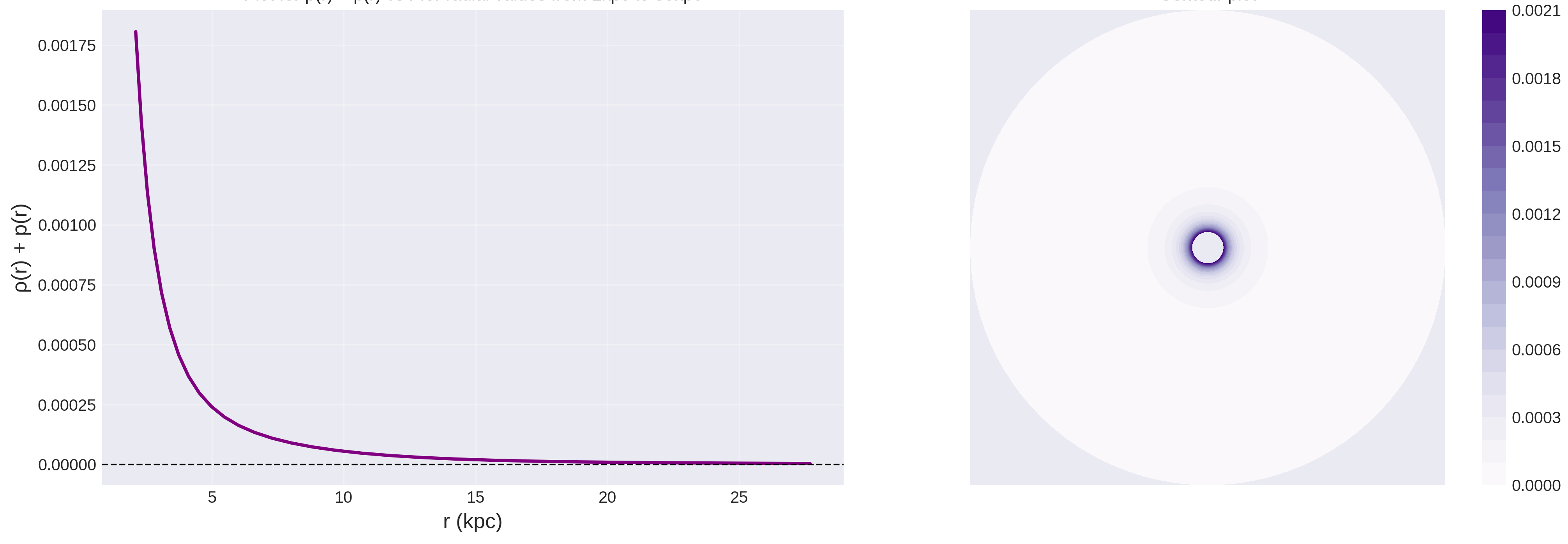}
    \caption{Plot for $\rho(r)+p(r)$ vs $r$ for radial values and its corresponding contour plot}
    \label{fig:placeholder}
\end{figure}

\subsection{Velocity of sound}

In the study of a galaxy modeled as a stellar object, the causality conditions require that the speed of sound do not exceed the speed of light. These constraints ensure physical viability:
\begin{align}\label{sound_eq}
0 \leq v_{\text{sound}}^2 = \frac{dp}{d\rho} \leq 1,
\end{align}
where $p$ is the pressure, and $\rho$ is the energy density. These inequalities guarantee that signals propagate at subluminal speeds.
\begin{figure}
    \centering
    \includegraphics[width=1\linewidth]{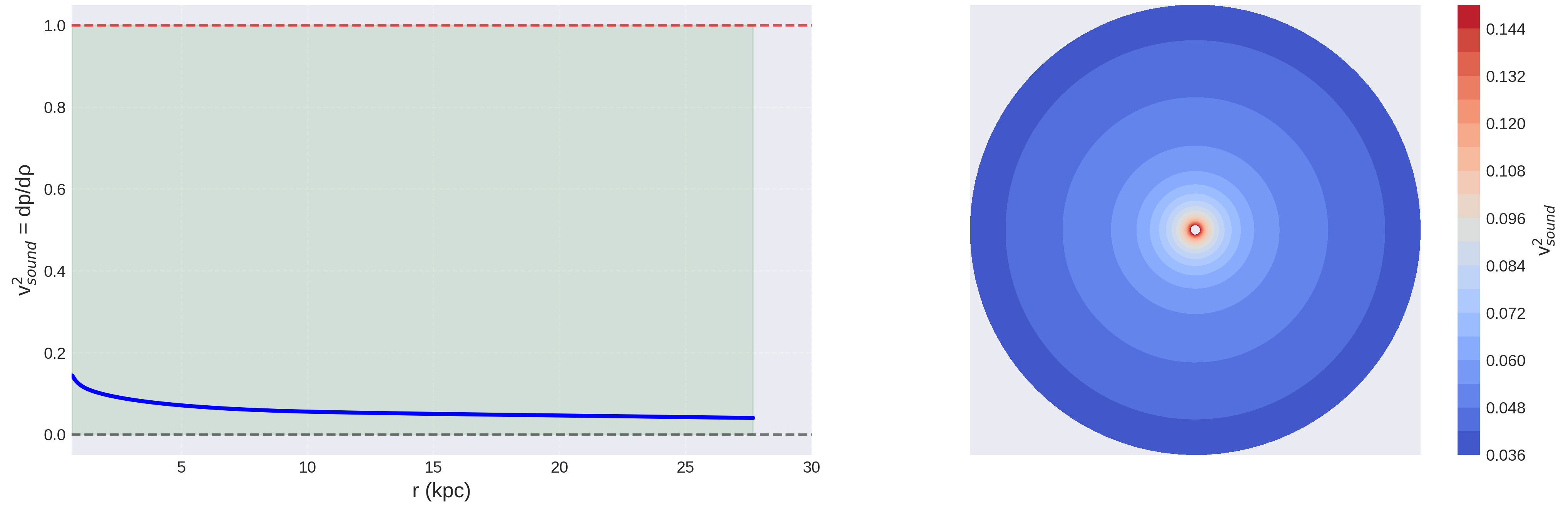}
    \caption{ Plot for the squared sound velocity vs $r$ and its corresponding contour plot}
    \label{fig:placeholder-1}
\end{figure}

From Fig. \ref{fig:placeholder-1}, we can clearly see that the plot of the squared velocity of sound indeed satisfies the constraint. Hence the causality conditions are also satisfied in our model.

\section{Stability of circular orbits}

Restricting to equatorial plane motion (\( \theta = \pi/2 \)) and defining the four-velocity \( U^\alpha = \frac{dx^\alpha}{d\tau} \), the normalization condition \( g_{\mu\nu} U^\mu U^\nu = -1 \) leads to a radial equation of the form
\[
\left(\frac{dr}{d\tau}\right)^2 = E^2 - V_{\text{eff}}(r),
\]
where the effective potential is given by
\[
V_{\text{eff}}(r) = \left(1 - \frac{2m(r)}{r} \right)\left(1 + \frac{J^2}{r^2}\right).
\]

Here \( E \) and \( J \) are the conserved relativistic energy and angular momentum per unit rest mass, defined by
\[
E = f(r)\dot{t}, \quad J = r^2\dot{\phi}.
\]

Circular orbits are defined by constant radius \( r = R \), so that
\[
\left. \frac{dr}{d\tau} \right|_{r = R} = 0 \quad \text{and} \quad \left. \frac{dV_{\text{eff}}}{dr} \right|_{r = R} = 0.
\]

Solving the above conditions yields the values of \( E \) and \( J \) corresponding to circular orbits. The stability of such orbits is determined by the sign of the second derivative of the effective potential:
\[
\text{Stable} \iff \left. \frac{d^2V_{\text{eff}}}{dr^2} \right|_{r = R} > 0, \qquad \text{Unstable} \iff \left. \frac{d^2V_{\text{eff}}}{dr^2} \right|_{r = R} < 0.
\]

Evaluating the second derivative explicitly, we find
\[
\begin{aligned}
\left. \frac{d^2 V_{\text{eff}}}{dr^2} \right|_{r = R} =\; &
\left( -\frac{2m''(R)}{R} + \frac{4m'(R)}{R^2} - \frac{4m(R)}{R^3} \right)
\left( 1 + \frac{J^2}{R^2} \right) \\
&+ \frac{8J^2}{R^3} \left( \frac{m'(R)}{R} - \frac{m(R)}{R^2} \right)  + \left( 1 - \frac{2m(R)}{R} \right) \left[ 
\frac{6J^2}{R^4} \right]
\end{aligned}
\]

The sign of this expression determines the nature of the circular orbits
\[
\left. \frac{d^2 V_{\text{eff}}}{dr^2} \right|_{r = R} > 0 \quad \Rightarrow \quad \text{Stable orbit,}
\]
\[
\left. \frac{d^2 V_{\text{eff}}}{dr^2} \right|_{r = R} < 0 \quad \Rightarrow \quad \text{Unstable orbit.}
\]

In particular, the effective potential admits a local minimum when
\begin{itemize}
    \item \( r > 2m(r) \) (i.e., outside any horizon),
    \item \( m'(r) > \dfrac{m(r)}{r} \).
\end{itemize}

These conditions ensure that the sum of terms contributing to \( \frac{d^2 V_{\text{eff}}}{dr^2} \) is positive, guaranteeing the stability of circular orbits in the given geometry.

\begin{figure}
    \centering
    \includegraphics[width=1\linewidth]{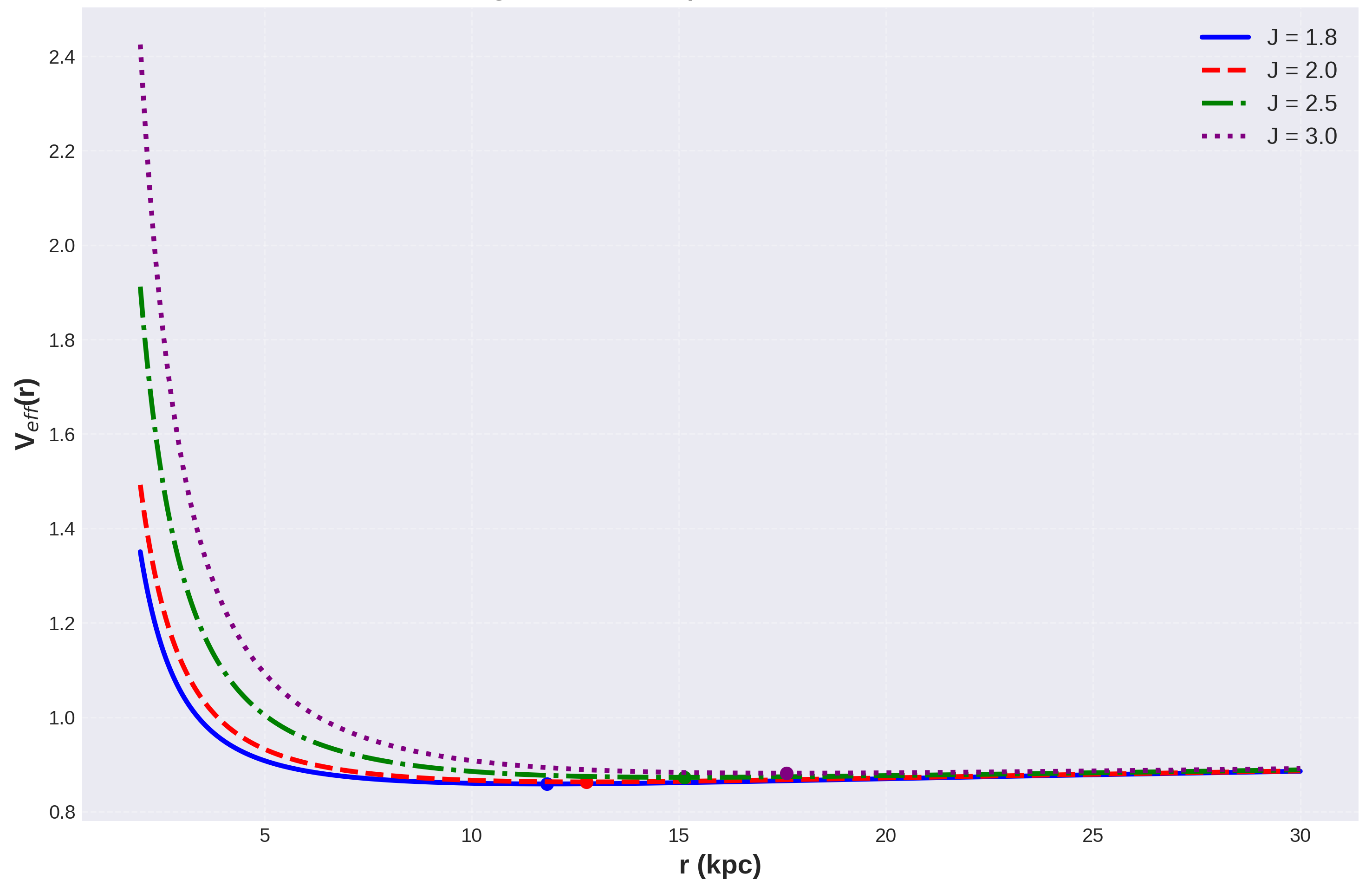}
     \caption{Plot for the effective potential $V_{\text{eff}}$ versus $r$ for angular momentum per unit mass $J = 1.8, 2, 2.5, 3$}
    \label{fig:placeholder}
\end{figure}

Here we see in each case that the effective potential has a minima which is supported by the positive value of the second-order derivative. Hence, they signify stable orbits.

\section{Attractive gravity from effective anisotropic stress}

We consider a test particle initially at rest in the equatorial plane (\( \theta = \pi/2 \)). The motion of the particle is governed by the geodesic equation:
\begin{equation}
    \frac{d^2 r}{d\tau^2} + \Gamma^r_{\mu\nu} \frac{dx^\mu}{d\tau} \frac{dx^\nu}{d\tau} = 0.
\end{equation}

At the point of release, the particle has no radial velocity or acceleration, i.e., \( \dot{r} = 0 \). Using the relevant Christoffel symbols, the radial geodesic equation simplifies to
\begin{equation}
    \left.\frac{d^2 r}{d\tau^2}\right|_{r = r_0} = -\Gamma^r_{tt} \left(\frac{dt}{d\tau}\right)^2 - \Gamma^r_{\phi\phi} \left(\frac{d\phi}{d\tau}\right)^2.
\end{equation}

The Christoffel symbols for the metric
\begin{equation}
    ds^2 = -f(r)\,dt^2 + \frac{dr^2}{1 - \dfrac{2m(r)}{r}} + r^2\left(d\theta^2 + \sin^2\theta\, d\phi^2\right)
\end{equation}
can be provided as follows:
\begin{equation}
    \Gamma^r_{tt} = \frac{f'(r)}{2} \left(1 - \frac{2m(r)}{r}\right), \quad
    \Gamma^r_{\phi\phi} = - r \left(1 - \frac{2m(r)}{r}\right).
\end{equation}

Substituting into the geodesic equation, we get
\begin{equation}
\left.\frac{d^2 r}{d\tau^2}\right|_{r = r_0} = -\left(1 - \frac{2m(r_0)}{r_0} \right)\left[\frac{f'(r_0)}{2} \left(\frac{dt}{d\tau}\right)^2 - r_0 \left(\frac{d\phi}{d\tau}\right)^2 \right].
\end{equation}

Since \( f'(r_0) > 0 \) and all squared quantities are positive, the right-hand side is {\it{negative}}, which means the radial acceleration is directed inward - confirming that gravity is attractive.

\subsection{Gravitational energy argument}

An alternative way to demonstrate the attractive nature of gravity is to compute the total gravitational energy between two radii \( r_1 \) and \( r_2 \). This is defined as
\begin{equation}
E_G = 4\pi \int_{r_1}^{r_2} \left[1 - \sqrt{g_{rr}(r)}\right] \rho(r)\, r^2\, dr,
\end{equation}
where the radial metric component is:
\begin{equation}
g_{rr}(r) = \frac{1}{1 - \dfrac{2m(r)}{r}}.
\end{equation}

So
\begin{equation}
\sqrt{g_{rr}(r)} = \left(1 - \frac{2m(r)}{r} \right)^{-1/2}.
\end{equation}

Hence
\begin{equation}
E_G = 4\pi \int_{r_1}^{r_2} \left[1 - \left(1 - \frac{2m(r)}{r} \right)^{-1/2}\right] \rho(r)\, r^2\, dr.
\end{equation}

We now substitute the expression for the energy density, which follows from the Einstein equation \( G^0_0 = 8\pi \rho \), yielding:
\begin{equation}
\rho(r) = \frac{1}{4\pi} \left[\frac{m'(r)}{r^2}\right].
\end{equation}

Substituting this into the integral
\begin{equation}\label{grav}
E_G = \int_{r_1}^{r_2} \left[1 - \left(1 - \frac{2m(r)}{r} \right)^{-1/2} \right] m'(r)\, dr.
\end{equation}

\begin{figure}
    \centering
    \includegraphics[width=1\linewidth]{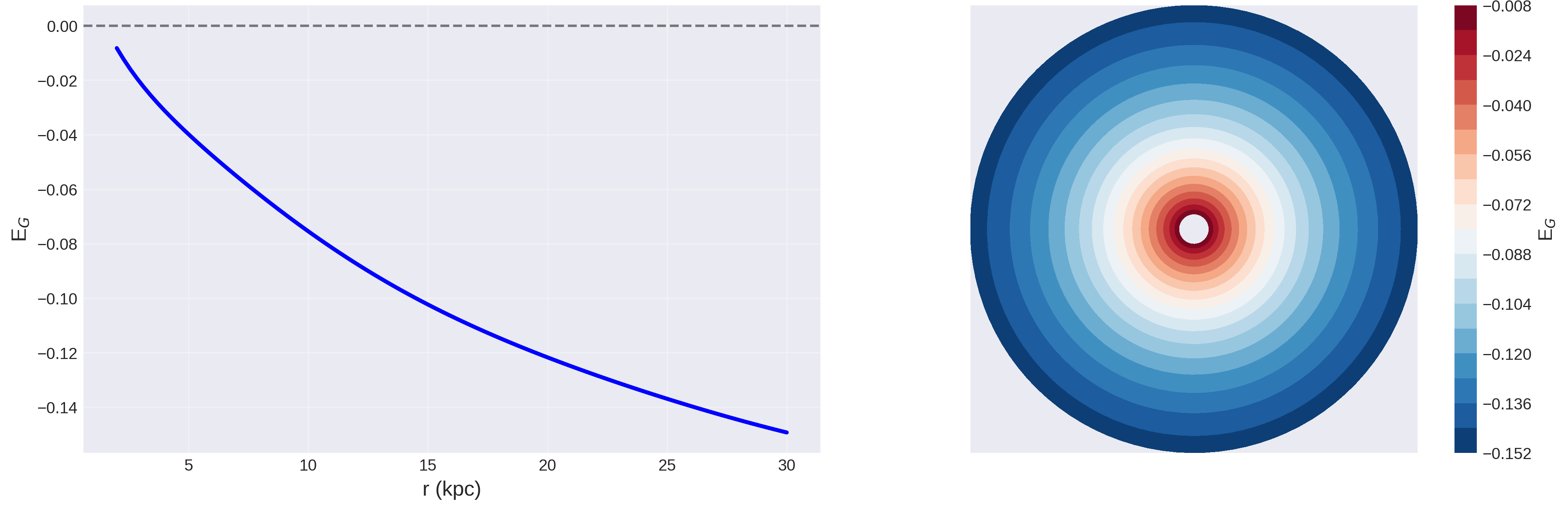}
    \caption{Plot of gravitational energy given by Eq. \eqref{grav} for fixed $r_1 = 1.5$ and variable $r_2$ and its corresponding contour plot}
    \label{fig:placeholder}
\end{figure}

From the above figure, it is evident that the gravitational energy is negative value, i.e.,
\begin{equation}
E_G < 0.
\end{equation}

\section{Explanation of the Plots}

\subsection{Rotation Curve Reconstruction (Figure 1)}

\textbf{Figure 1:} The machine-learning-reconstructed Milky Way rotation curve shows strong agreement with the 73 observational data points spanning 0.1--95.56 kpc. The Ridge regression model with $R^2 = 0.9824$ and RMSE $= 3.75$ km s$^{-1}$ successfully captures the rapid inner rise (0.1--0.4 kpc), extended flat plateau (0.4--15 kpc), secondary peak ($\sim$15 kpc), and gradual outer decline (15--95.56 kpc). The resulting analytical solution is smooth and differentiable, allowing for a direct embedding into Einstein’s field equations. The current reconstruction method does not rely on any specific conventional shape of the halo profile, but rather a flexible and physically based method with the help of data-driven regularization to finally obtain the analytical shape.

\subsection{Model Comparison (Figures 2--6)}

\textbf{Figure 2:} Comparative analysis on logarithmic scale reveals the Ridge regression model's superiority over conventional NFW and Burkert profiles. The polynomial parametrization tracks observational data closely across the entire radial domain, while NFW systematically underestimates velocities in intermediate regions ($\chi^2_{\text{red}} = 11.05$) and Burkert shows marginal improvement ($\chi^2_{\text{red}} = 7.44$). The Ridge model achieves $\chi^2_{\text{red}} = 0.98$, confirming that data-driven reconstruction without rigid parametric assumptions provides superior observational fidelity.

\textbf{Figures 3--5:} Individual model comparisons with 1$\sigma$ and 2$\sigma$ confidence intervals quantify predictive reliability. The Ridge model maintains narrow uncertainty bands ($\sigma \approx 8.71$ km s$^{-1}$) throughout, while NFW exhibits large scatter ($\sigma \approx 4.95 \pm 2.20$ km s$^{-1}$, $R^2 = 0.64$) and Burkert shows moderate dispersion ($\sigma \approx 3.31 \pm 1.92$ km s$^{-1}$, $R^2 = 0.79$). The machine learning approach outperforms traditional halo models in both accuracy and stability.

\textbf{Figure 6:} Radial dependence of prediction uncertainty $\sigma(r)$ demonstrates the Ridge model's consistent low uncertainty across all radii, contrasting sharply with the elevated and variable uncertainties of NFW and Burkert models, particularly in outer halo regions where parametric forms struggle to adapt to observational complexity.

\subsection{Mass Function (Figure 7)}

\textbf{Figure 7:} The mass function $m(r)$ exhibits smooth monotonic growth throughout 2--30 kpc, confirming a physically acceptable extended halo without mass shells, discontinuities, or pathological behavior. The contour plot reveals perfect spherical symmetry consistent with the static, spherically symmetric metric assumption. The continuous mass accretion with radius is essential for sustaining the observed rotation curve behavior, with $m(r)$ derived self-consistently from the machine-learning velocity profile through Einstein's field equations.

\subsection{Density Profile (Figure 8)}

\textbf{Figure 8:} The tangential pressure $p(r)$ remains positive throughout and decreases monotonically outward, with maximum values near the galactic center where gravitational binding is strongest. The contour demonstrates spherically symmetric stress distribution characteristic of anisotropic fluid configurations.  With ($p_r = 0$) and ($p_t \neq 0$), the stress-energy profile mimics dark matter gravity through purely geometric means, without the need for exotic matter.

\subsection{Tangential Pressure (Figure 9)}

\textbf{Figure 9:} The equation of state parameter $\omega = p/\rho \in [0.035, 0.084]$ remains small and positive throughout 2--30 kpc, indicating matter-dominated behavior. The decreasing trend with radius reflects progressive softening of the effective fluid equation of state. All values satisfy $\omega \ll 1/3$, ruling out radiation-dominated regimes ($\omega = 1/3$) and dark energy behavior ($\omega < 0$ or $\omega \approx -1$). This circular contour represents the assumed symmetry of the spherical model.

\subsection{Equation of State (Figure 10)}

\textbf{Figure 10:} The dominant energy condition $\rho - p \geq 0$ is satisfied everywhere, ensuring energy density exceeds pressure magnitude throughout the halo. All values are positive throughout the domain, showing consistency with non-exotic matter distributions. The smooth radial decay and spherically symmetric contour validate the model's consistency with standard matter distributions.

\subsection{Energy Conditions (Figures 11--13)}

\textbf{Figure 11:} The strong energy condition $\rho + 2p \geq 0$ holds across the entire domain, guaranteeing attractive gravity in the classical sense. The monotonically decreasing profile with positive values throughout indicates weakening effective mass-energy density at larger radii, consistent with expected galactic halo structure without exotic physics.

\textbf{Figure 12:} The null/weak energy condition $\rho + p \geq 0$ is fully satisfied, confirming energy flows along timelike or null geodesics. Positive values establish that the reconstructed spacetime does not require negative energy densities, maintaining compatibility with standard matter while reproducing dark-matter-like gravitational effects through anisotropic pressure.

\textbf{Figure 13:} The quantity $\rho(r) + p(r)$ remains positive throughout the considered radial range, confirming the validity of the null and weak energy conditions. The profile shows a smooth monotonic decrease with increasing radius, indicating a gradual reduction in the effective energy density contribution. The contour plot exhibits clear spherical symmetry, consistent with the underlying static and spherically symmetric spacetime geometry. The absence of any negative regions ensures that no exotic matter is required, and the physical behavior remains well within the bounds of standard relativistic matter distributions.

\subsection{Causality (Figure 14)}

\textbf{Figure 14:} The squared sound speed $v_s^2 = dp/d\rho \in [0.04, 0.14]$ satisfies the causality constraint $0 \leq v_s^2 \leq 1$ throughout the 0.65--30 kpc range. Perturbations propagate at $\sim$20--38\% of light speed, ensuring subluminal signal transmission and maintaining consistency with special relativity. The contour plot demonstrates radially varying but everywhere physical sound speed distribution, with no superluminal or acausal behavior. This validates the model's dynamical stability and physical consistency.

\subsection{Orbital Stability (Figure 15)}

\textbf{Figure 15:} The effective potential $V_{\text{eff}}(r)$ exhibits clear minima for all tested angular momentum values ($J = 1.8, 2.0, 2.5, 3.0$ kpc$^2$), confirming stable circular orbits throughout the galactic disk and inner halo. Higher angular momentum generates stronger centrifugal barriers, shifting orbital radii outward: $J = 1.8$ yields $r_{\text{min}} \approx 11.8$ kpc, while $J = 3.0$ gives $r_{\text{min}} \approx 17.6$ kpc. The positive second derivatives ($d^2V_{\text{eff}}/dr^2 > 0$) at all minima guarantee orbital stability against small perturbations, consistent with long-lived stellar and gas dynamics observed in spiral galaxies like the Milky Way.

\subsection{Gravitational Energy (Figure 16)}

\textbf{Figure 16:} The gravitational energy $E_G < 0$ throughout the integration domain from $r_1 = 1.5$ kpc to variable $r_2 \leq 30$ kpc confirms attractive gravity. The increasingly negative values with radius ($E_G \in [-0.149, -0.008]$) reflect deeper gravitational binding as more mass is enclosed, characteristic of bound gravitational systems. The contour plot demonstrates spherically symmetric energy distribution with more negative values (stronger binding) at larger radii. This independent verification through the gravitational energy integral complements the geodesic deviation analysis, establishing that the machine-learning-reconstructed spacetime describes an attractive gravitational field fully consistent with observed galactic structure and dynamics.

\subsection{Overall Physical Consistency}

Overall, Figs. 7--16 show that the machine learning model-derived velocity profile, when implemented in Einstein’s equations, produces a physically viable space time geometry. All energy conditions are satisfied (Figures 11--13), causality is preserved (Figure 14), circular orbits are stable (Figure 15), and gravity is attractive (Figure 16). The model successfully reproduces dark-matter-like gravitational effects through geometric anisotropic stress without invoking exotic matter, phantom fields, or violations of classical energy conditions, providing a self-consistent and observationally validated description of Milky Way galactic dynamics.

\section{Discussion and Conclusion}

In this work, we have presented a machine-learning-guided framework for reconstructing the Milky Way rotation curve and examining its implications within a relativistic spacetime description. By adopting a physically motivated basis of functions and systematically comparing Ridge regression, LASSO regression, and a feed-forward neural network, we demonstrate that Ridge regression offers the most effective balance between accuracy, robustness, and analytical transparency. This choice is crucial, as it yields an explicit velocity profile that can be directly embedded into Einstein’s field equations. Unlike conventional parametric dark matter halo models, which impose restrictive functional forms on the velocity profile, the present approach allows the observational data to guide the reconstruction while maintaining smoothness and differentiability across the entire radial domain. 

A few salient features of the present study are as follows:

(i) The resulting machine-learning-derived rotation curve reproduces the observed kinematic features of the Milky Way with high fidelity and significantly outperforms standard NFW and Burkert models in terms of goodness-of-fit and predictive accuracy. 

(ii) Embedding the reconstructed velocity profile into a static and spherically symmetric spacetime leads to a self-consistent determination of the redshift and mass functions. 

(iii) The effective matter content supporting the geometry is described by an anisotropic fluid with vanishing radial pressure. 

(iv) A detailed physical analysis shows that the energy density remains positive, the tangential pressure is well behaved, and all classical energy conditions are satisfied throughout the galactic halo. Moreover, the equation of state remains within physically acceptable bounds, and the causality condition is preserved, with the speed of sound remaining subluminal over the full radial range.

(v) The stability of the spacetime is further confirmed through an analysis of circular geodesics, which demonstrates the existence of stable orbits, consistent with long-lived galactic structures. 

(vi) Independent confirmation of the attractive nature of gravity is obtained from both the geodesic acceleration and gravitational energy considerations, reinforcing the internal consistency and physical viability of the reconstructed geometry.

Overall, the present study establishes that machine-learning-assisted analytical reconstruction of galactic rotation curves provides a transparent and physically consistent bridge between observational data and relativistic gravitational modeling. By avoiding rigid parametric assumptions while preserving interpretability, the framework developed here offers a flexible alternative for studying galactic halos. 

Furthermore, a few important physically demanding points are to be discussed here under special comments as: 

(1) While the Milky Way is a rotating disk galaxy, spherical symmetry provides a well-established approximation for large-scale halo dynamics beyond the inner disk region. Observational studies indicate that the dark matter halo is approximately spherical or mildly oblate at radii greater than ~5 kpc. Since the present analysis focuses on the effective gravitational potential rather than detailed baryonic disk structure, the static spherically symmetric approximation captures the dominant radial dynamics relevant for rotation curve reconstruction. Extension to axisymmetric or rotating spacetimes (e.g., Kerr-like metrics) remains an interesting direction for future work, building on the redshift-function framework established in~\cite{rahaman2026observationally}.

(2) This framework does not replace dark matter but provides a complementary, data-driven reconstruction of the effective gravitational mass distribution. The anisotropic fluid represents the stress-energy content of the halo, just like dark matter, but with the NFW and Burkert profiles. This method does not impose stringent constraints on the halo profile, reconstructing the profile directly from the data while still being consistent with general relativity.

(3) The reconstructed mass distribution can be compared with that predicted by the $\Lambda$CDM model of cosmology. In the $\Lambda$CDM model, the profile of the halo is usually given by the NFW  model, in which the density decreases towards the outer regions of the halo. The reconstructed profile in this study has a similar general profile, producing a flat rotation curve and a gradual decline in the outer regions. Although no profile is specified, the mass distribution is generally consistent with the predictions of the $\Lambda$CDM model on large scales.                      

In principle, in our next work we would like to conduct a dedicated study regarding the observational constraints on the model parameters following our previous work \cite{rahaman2026observationally,Sanyal2026mnras} as that has been out of scope here due to special focus on the ML related applications.

\section*{CRediT authorship contribution statement}
Aritra Sanyal: Visualization, Software, Writing – original draft. Swapan Das: Methodology, Writing – original draft. Farook Rahaman: Conceptualization, Methodology, Formal analysis, Validation. Saibal Ray: Conceptualization, Writing – review \& editing, Supervision.

\section*{Declaration of competing interest}
The authors declare that they have no known competing financial interests or personal relationships that could have appeared to influence the work reported in this paper.

\section*{Acknowledgment}
FR, SR, and AS are thankful to the Inter-University Centre for Astronomy and Astrophysics (IUCAA), Pune, India, for their support.  FR and AS also gratefully acknowledge academic support from Jadavpur University (JU). FR thanks ANRF, SERB, Government of India, for their support.

\section*{Data availability}
Generated dataset is available in the Table of the manuscript.

\section*{ORCID}

Aritra Sanyal: 0009-0003-0383-8335

Swapan Das: 0009-0005-3773-2189

Farook Rahaman: 0000-0003-0594-4783

Saibal Ray: 0000-0002-5909-0544

\appendix

\section{Rotation Curve Dataset for the Milky Way}
\label{app:RCdata}

\begin{longtable}{ccc}
\caption{Rotation curve of the Milky Way used in Figure~1.}
\label{tab:RC_MilkyWay}\\
\hline\hline
Radius (kpc) & $V_{\rm rot}$ (km s$^{-1}$) & Standard Deviation (km s$^{-1}$) \\
\hline
\endfirsthead

\multicolumn{3}{c}%
{{\bfseries Table~\thetable\ continued from previous page}}\\
\hline
Radius (kpc) & $V_{\rm rot}$ (km s$^{-1}$) & Standard Deviation (km s$^{-1}$) \\
\hline
\endhead

\hline
\multicolumn{3}{r}{{Continued on next page}}\\
\endfoot

\hline\hline
\endlastfoot

0.100 & 144.9 & 3.7 \\
0.110 & 147.4 & 4.2 \\
0.121 & 150.4 & 4.8 \\
0.133 & 153.8 & 6.1 \\
0.146 & 158.9 & 10.3 \\
0.161 & 167.4 & 16.1 \\
0.177 & 180.1 & 22.4 \\
0.195 & 196.6 & 27.1 \\
0.214 & 213.6 & 26.9 \\
0.236 & 227.8 & 22.7 \\
0.259 & 237.9 & 17.0 \\
0.285 & 244.4 & 11.8 \\
0.314 & 248.2 & 7.6 \\
0.345 & 250.2 & 4.7 \\
0.380 & 251.0 & 2.9 \\
0.418 & 250.7 & 2.1 \\
0.459 & 249.7 & 2.3 \\
0.505 & 248.0 & 2.9 \\
0.556 & 245.9 & 3.7 \\
0.612 & 243.2 & 4.6 \\
0.673 & 239.8 & 5.7 \\
0.740 & 235.8 & 6.4 \\
0.814 & 231.7 & 6.5 \\
0.895 & 227.8 & 6.0 \\
0.985 & 224.5 & 5.2 \\
1.083 & 221.7 & 4.5 \\
1.192 & 219.1 & 4.0 \\
1.311 & 216.8 & 3.7 \\
1.442 & 214.7 & 3.4 \\
1.586 & 212.7 & 3.1 \\
1.745 & 210.9 & 2.8 \\
1.919 & 209.5 & 2.3 \\
2.111 & 208.5 & 1.8 \\
2.323 & 208.2 & 1.6 \\
2.555 & 208.9 & 2.2 \\
2.810 & 210.7 & 3.6 \\
3.091 & 213.4 & 4.8 \\
3.400 & 217.2 & 5.9 \\
3.740 & 222.0 & 6.6 \\
4.114 & 226.6 & 5.7 \\
4.526 & 229.5 & 4.4 \\
4.979 & 231.6 & 4.3 \\
5.476 & 234.1 & 5.3 \\
6.024 & 237.2 & 5.7 \\
6.626 & 239.5 & 5.0 \\
7.289 & 240.1 & 4.1 \\
8.018 & 239.0 & 4.4 \\
8.820 & 236.7 & 5.4 \\
9.702 & 234.5 & 6.0 \\
10.672 & 234.2 & 7.1 \\
11.739 & 237.1 & 9.8 \\
12.913 & 242.8 & 12.4 \\
14.204 & 248.5 & 13.3 \\
15.625 & 249.7 & 14.8 \\
17.187 & 246.2 & 17.4 \\
18.906 & 243.3 & 18.3 \\
20.797 & 243.9 & 17.5 \\
22.876 & 245.6 & 15.6 \\
25.164 & 243.7 & 15.2 \\
27.680 & 237.3 & 16.1 \\
30.448 & 229.6 & 15.5 \\
33.493 & 222.5 & 14.1 \\
36.842 & 215.0 & 14.0 \\
40.527 & 207.1 & 13.8 \\
44.579 & 200.3 & 12.7 \\
49.037 & 194.7 & 11.9 \\
53.941 & 189.8 & 11.3 \\
59.335 & 186.2 & 10.4 \\
65.268 & 184.7 & 9.6 \\
71.795 & 183.9 & 9.3 \\
78.975 & 181.4 & 11.0 \\
86.872 & 175.5 & 14.6 \\
95.560 & 167.7 & 16.3 \\
\end{longtable}

\section{Ridge Regression Model}

The Ridge regression model has a high predictive potential, with a cross-validated coefficient of determination given by
\begin{equation}
R^2_{\mathrm{CV}} = 0.9824 \pm 0.0064,
\end{equation}
and a root-mean-square error:
\begin{equation}
\mathrm{RMSE}_{\mathrm{CV}} = 3.75 \pm 1.17 \; \mathrm{km\,s^{-1}}.
\end{equation}

The low fold-to-fold variance indicates statistical stability and absence of overfitting.

In contrast, the predictive potential of the NFW halo model is much lower:
\begin{equation}
R^2 = 0.643 \pm 0.032,
\end{equation}
\begin{equation}
\mathrm{RMSE} = 16.88 \pm 3.84 \; \mathrm{km\,s^{-1}},
\end{equation}
indicating poorer agreement with observational data.

The Burkert profile has a slightly better predictive potential than the NFW model but still performs worse than the data-driven reconstruction:
\begin{equation}
R^2_{\mathrm{CV}} = 0.792 \pm 0.041,
\end{equation}
\begin{equation}
\mathrm{RMSE}_{\mathrm{CV}} = 12.90 \pm 5.31 \; \mathrm{km\,s^{-1}}.
\end{equation}

\end{document}